\documentclass[journal]{IEEEtran}
\usepackage{graphicx,amsmath,booktabs,tabularx,float,url,placeins}

\usepackage[caption=false,font=footnotesize]{subfig}

\graphicspath{{Role_Paper_People_s_Music/}}

\begin{document}
\title{From Textural Counterpoint to Feature Encoding: A Multi-Dimensional Machine Representation Study of Haydn's \textit{The Lark} Integrating Electroacoustic Analysis}

\author{Yakun~Liu, 
        Zhiyu~Jin, 
        Hai~Luan, 
        Dong~Liu, 
        and~Xiaonan~Li
\thanks{Corresponding author: Xiaonan Li.}%
\thanks{Yakun Liu, Dong Liu, and Xiaonan Li are with the Department of Composition, Shenyang Conservatory of Music, Shenyang 110818, Liaoning, China.}%
\thanks{Zhiyu Jin is with the Department of Musicology, Shenyang Conservatory of Music, Shenyang 110818, Liaoning, China.}%
\thanks{Hai Luan is with the Education Information Center, Shenyang Conservatory of Music, Shenyang 110818, Liaoning, China.}%
}

\maketitle

\begin{abstract}
Chamber music, as a highly precise multi-part interactive system, contains a logic of "role assignment and dynamic interaction" that provides an extremely valuable blueprint for exploring human-computer collaborative composition paradigms. Addressing the lack of role perception capabilities in existing deep music generation models during polyphonic interactions, this paper conducts an interdisciplinary analysis of Haydn's String Quartet in D Major, \textit{The Lark} (Op. 64, No. 5). We propose a novel research path: "Classical Morphology Qualitative Analysis—Electroacoustic Quantitative Measurement—Machine Representation Reconstruction." The study first utilizes auditory analysis to dissect the counterpoint morphology of the leading voice and the underlying groove in the first movement. Subsequently, it introduces spectrum and dynamic feature analysis tools from a Digital Audio Workstation (DAW) to translate subjective auditory perception into objective, measurable physical parameters. Building on this, the paper introduces a fundamentally new approach to low-level computer feature extraction: completely abandoning the traditional mechanical quantization grid, introducing Event-based Timestamps to record the duration of micro-timing, and transforming acoustic features into an independent "Role-Aware Encoding" as an aesthetic heuristic mechanism (a phenomenological anchor). This study not only completes the logical loop spanning classical analysis, electronic music mapping, and AI symbolic generation but also establishes a profound theoretical foundation—from the perspectives of interactive aesthetics and media philosophy—for constructing human-computer collaborative music systems imbued with "social attributes" and "otherness awareness."
\end{abstract}


\section{Introduction}
\subsection{Research Background and the Necessity of an Interdisciplinary Perspective}
Chamber music, particularly the string quartet, functions as a conductor-less, highly precise multi-part interactive system that demands an exceptional degree of tacit "listening" and "responding" among players \cite{ref1}. This musical form, relying purely on communication across multiple roles, provides a highly valuable closed research blueprint for exploring human-computer collaborative composition paradigms. As a core genre established and brought to its zenith during the Classical period, its primary musical tension stems not merely from stacked harmonies, but from highly explicit and fluid role assignments and dynamic handovers.

However, as artificial intelligence advances deeper into symbolic music generation \cite{ref2, ref3}, the rapid surge of technicalism has masked fatal flaws in the underlying logic of these models. When handling multi-part music, mainstream deep learning models often reduce music to flat, non-interfering arrays of notes. Traditional algorithms heavily rely on a fixed grid for forced spatial quantization. This purely engineering-driven approach brutally obliterates the micro-timing elasticity (rubato) crucial to chamber music \cite{ref4, ref5} and completely deprives the model of any role perception capability. To break this bottleneck in machine generation, an interdisciplinary mapping pathway must be established: classical musicology's "qualitative analysis" is needed to provide an a priori blueprint of organizational logic, while electronic music engineering's "quantitative measurement" is required to provide the physical texture of sound morphologies. Ultimately, these must coalesce to reconstruct "data representation" at the low-level inputs of machine learning.

\subsection{Core Contributions of This Paper}
Addressing these pain points, this paper proposes an intersecting analysis and reconstruction framework. First, it outlines the textural separation characteristics in the first movement of \textit{The Lark} from a classical auditory perspective. Second, it introduces modern electronic music acoustic measurement tools (such as spectrum, transient, and RMS analyzers) to physically profile live performance recordings. Finally, relying on the informational neutralization provided by acoustic data, we design a novel composite feature vector for deep temporal models that includes "Event-based Timestamps" and "Role-Aware Encoding." This study steps outside the framework of purely symbolic generation. By cross-modally extracting the interaction logic of voice roles in chamber music, it establishes the underlying representational cornerstone and philosophical discourse for building human-computer collaborative music systems.

\section{Classical Perspective: Auditory Qualitative Analysis and Textural Separation}
Composed in 1790 by the "Father of the String Quartet," Joseph Haydn, the String Quartet in D Major, \textit{The Lark} (Op. 64, No. 5) \cite{ref6, ref7}, is renowned for its exquisite textural layout. Its first movement exhibits highly representative textural separation characteristics: the first violin soars in the high register like a lark, while the lower voices maintain a highly mechanical and groovy foundation. This "dominant-counterpoint-support" interactive logic makes it an excellent paradigm for reconstructing machine representations.

\subsection{The "Lark" Motif and Voice Role Counterpoint}
The exposition of the first movement (Allegro moderato) of \textit{The Lark} demonstrates the ultimate state of textural separation as homophony matured in the Classical period \cite{ref23}. In the opening measures (m. 1-8), the cello, viola, and second violin form an extremely solid low- and mid-register cluster. Using highly granular staccato and syncopated figures, they establish a driving 2/2 (Alla breve) marching gait in the lower register, precisely acting as the "groove foundation" and "harmonic support framework."

Against this solid and restrained acoustic backdrop, the first violin spirals in with the bright timbre of its extreme high register (E string). Its broad intervallic leaps and continuous, singing long tones create a stark auditory contrast to the short staccatos of the lower three voices. Here, the first violin acts as the "leading melodic narrator." This comprehensive opposition in textural thickness, articulation, and register establishes the most fundamental interactive state machine within the music \cite{ref24}.

\begin{figure}[t]
    \centering
    \includegraphics[width=\linewidth]{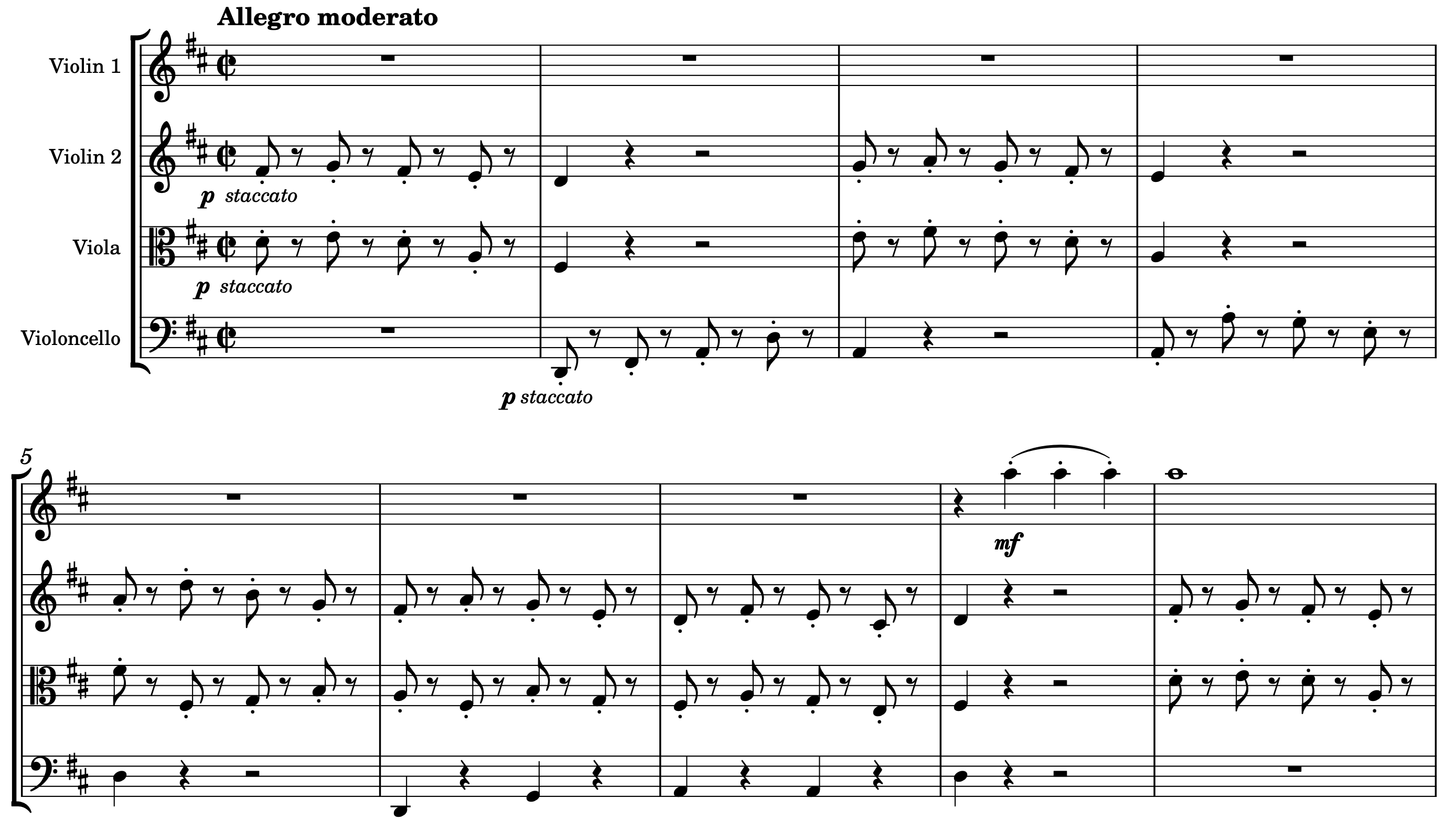}
    \caption{Musical score showing textural separation and auditory role counterpoint in the opening of \textit{The Lark}'s first movement.}
    \label{fig:lark_score}
\end{figure}

\subsection{Dynamic Switching and Yielding of the Role State Machine}
If the exposition establishes static role definitions, the development section fully highlights the fluidity of these role relationships. For instance, in the core development area (mm. 72-84), the first violin's dominance is no longer absolute; the core "lark" motif begins a dramatic, brief yielding to the mid-low voices (particularly the cello). The instruments rapidly switch among "leading core," "harmonic filler," and "contrapuntal response" through frequent imitation and sequence of short motifs. This highly precise "multi-role state machine," full of negotiation and compromise, is a tacit understanding built by human musicians through long-term ensemble training. If this fluid logic cannot be abstracted and captured, AI generation algorithms will forever remain stuck at one-way "melody manufacturing" and will be unable to learn true chamber music interaction logic.

\begin{figure}[t]
    \centering
    \includegraphics[width=\linewidth]{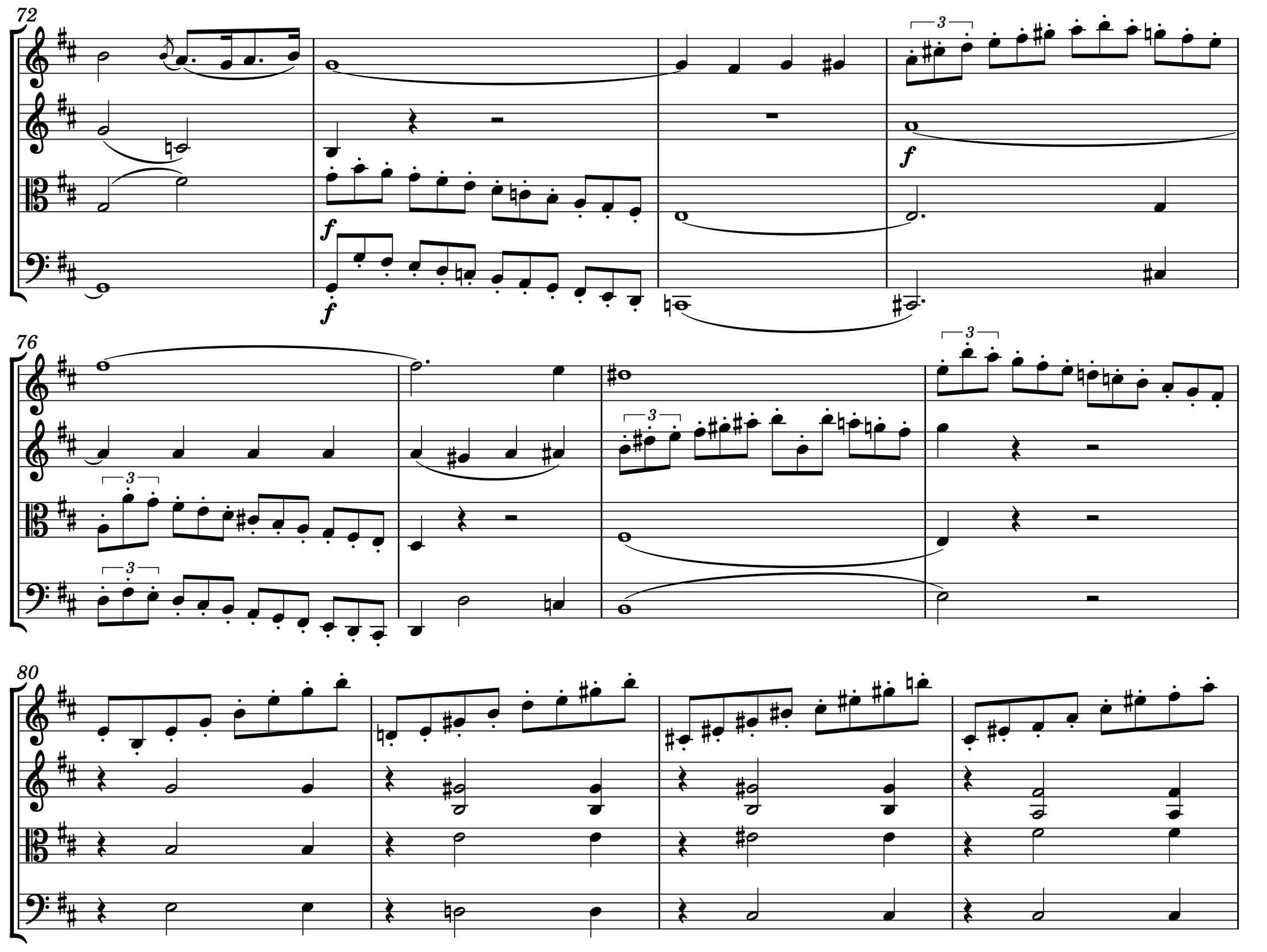}
    \caption{Musical score showing the dynamic switching of the role state machine during voice transitions in \textit{The Lark}'s first movement.}
    \label{fig:lark_analysis}
\end{figure}

\section{Electroacoustic Perspective: Physical Profiling of Phenomenological Anchors}
Traditional symbolic score data can only provide discrete pitch and duration information; machine learning models cannot derive the acoustic dimensional differences between leading and accompanying voices in physical space from symbolic sequences alone. To enable algorithms to "understand" role interactions in a classical context, this study introduces a Digital Audio Workstation (DAW) \cite{ref8} to conduct an electroacoustic analysis of the first eight measures of \textit{The Lark}, corroborated by objective "phenomenological anchors."

\subsection{Acoustic Metaphors of Frequency Band Energy Distribution}
By observing the peaks of the transcribed audio waveforms using the PAZ Frequency plugin in a high-precision Spectrum Analyzer, we found that the roles defined in the score possess natural frequency domain isolation in real physical space. As shown in Fig. \ref{fig:spectrum_all} (full independent spectral profiles of all four voices are detailed in Appendix A), the primary energy peak of the first violin's leading motif (approx. 903Hz) and its bright harmonics are highly concentrated in the mid-high frequency band. Conversely, the acoustic energy of the lower cello sinks and stabilizes in the ultra-low 80–300Hz foundational band. This physical isolation in frequency bands effectively reduces frequency masking between voices \cite{ref8}. It is no longer reliant purely on a composer's intuition, but rather objectively confirms the textural layering logic of "dominant melody" versus "underlying foundation support" using acoustic measurement data.

\begin{figure}[t]
    \centering
    \includegraphics[width=\linewidth]{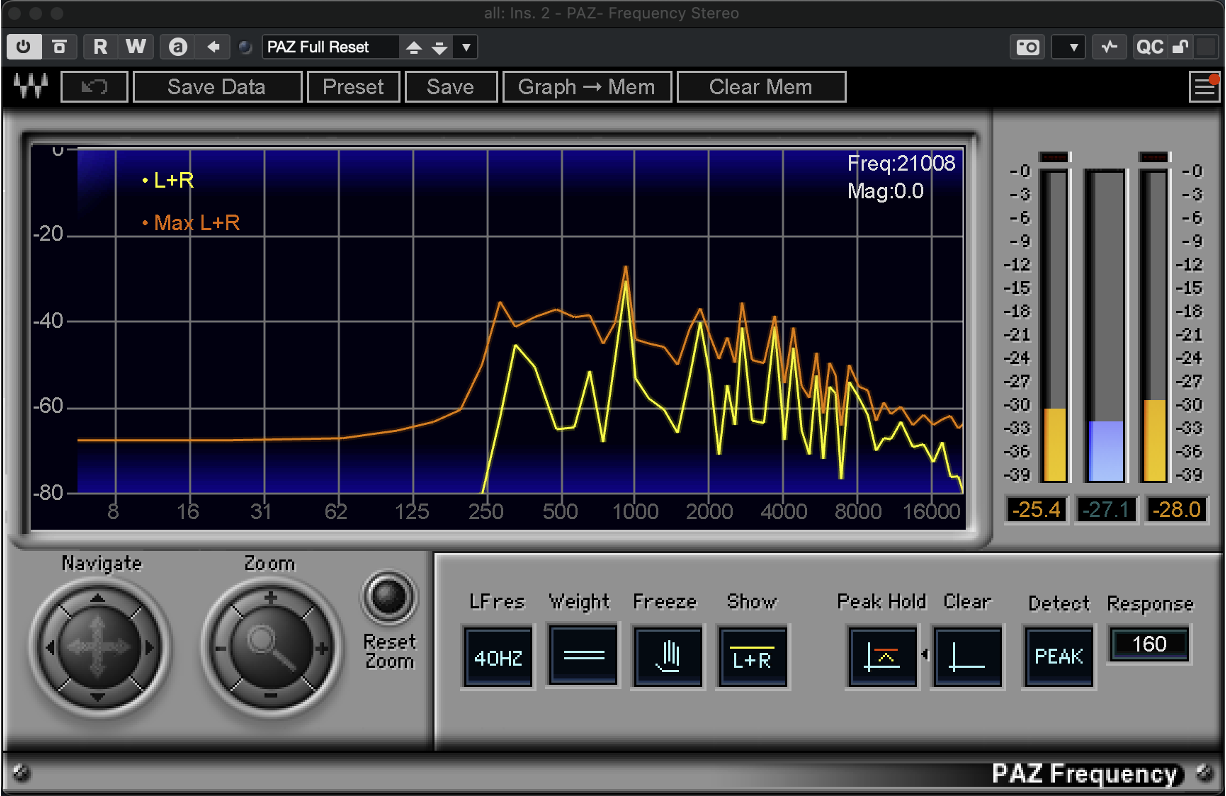}
    \caption{Core comparison of frequency domain layering for the four quartet voices (reflecting acoustic dimension role isolation).}
    \label{fig:spectrum_all}
\end{figure}

\subsection{Time-Domain Mapping of Transient Pulses and Articulation}
Beyond frequency distribution, classical contrasts such as "legato" and "staccato" display distinctly different acoustic fingerprints in the time domain. Relying on the binary opposition of transient density, vague and abstract classical auditory concepts can be translated into time-domain acoustic features that can be parsed directly by low-level computer systems. Using transient Hitpoints detection on the waveforms \cite{ref9} (see Fig. \ref{fig:hitpoints_main}; full four-voice comparisons in Appendix B, typical ADSR envelopes in Appendix C), the underlying cello accompaniment appears as dense, highly aggressive transient pulses with extremely fast and independent attack triggers. This perfectly replicates the "highly granular staccato" mentioned in Section II on a physical waveform, constituting a precise, machine-like metronome effect. In contrast, the first violin's sweeping melodic line shows highly glued notes, where transient detection barely registers any dense energy spikes, accurately corresponding to its "singing legato" designation.

\begin{figure}[t]
    \centering
    \subfloat[First Violin Transients (Legato lines, extremely low density)]{\includegraphics[width=\linewidth]{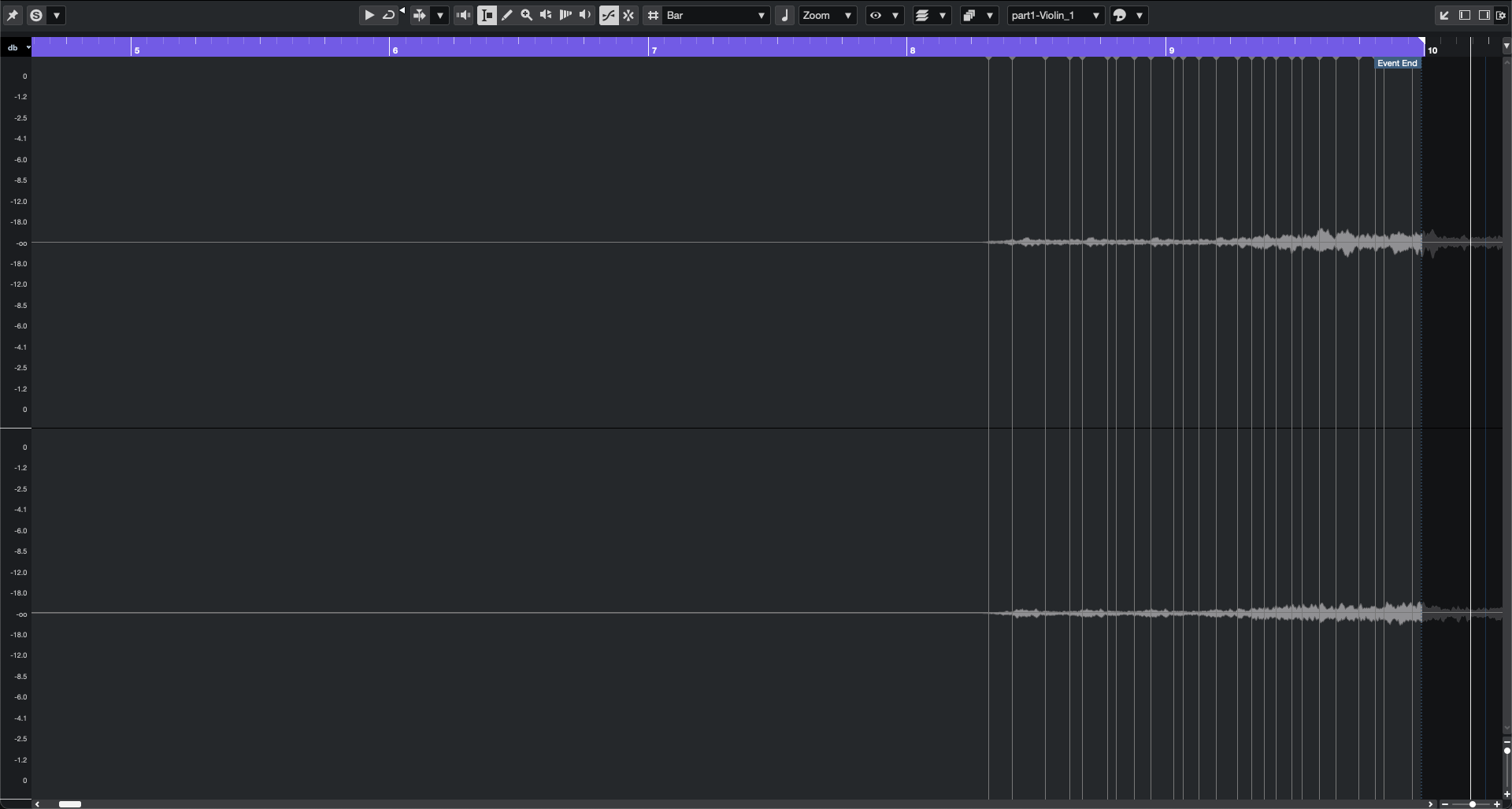}} \\
    \vspace{0.2cm}
    \subfloat[Cello Transients (Bass groove, high-density uniform pulses)]{\includegraphics[width=\linewidth]{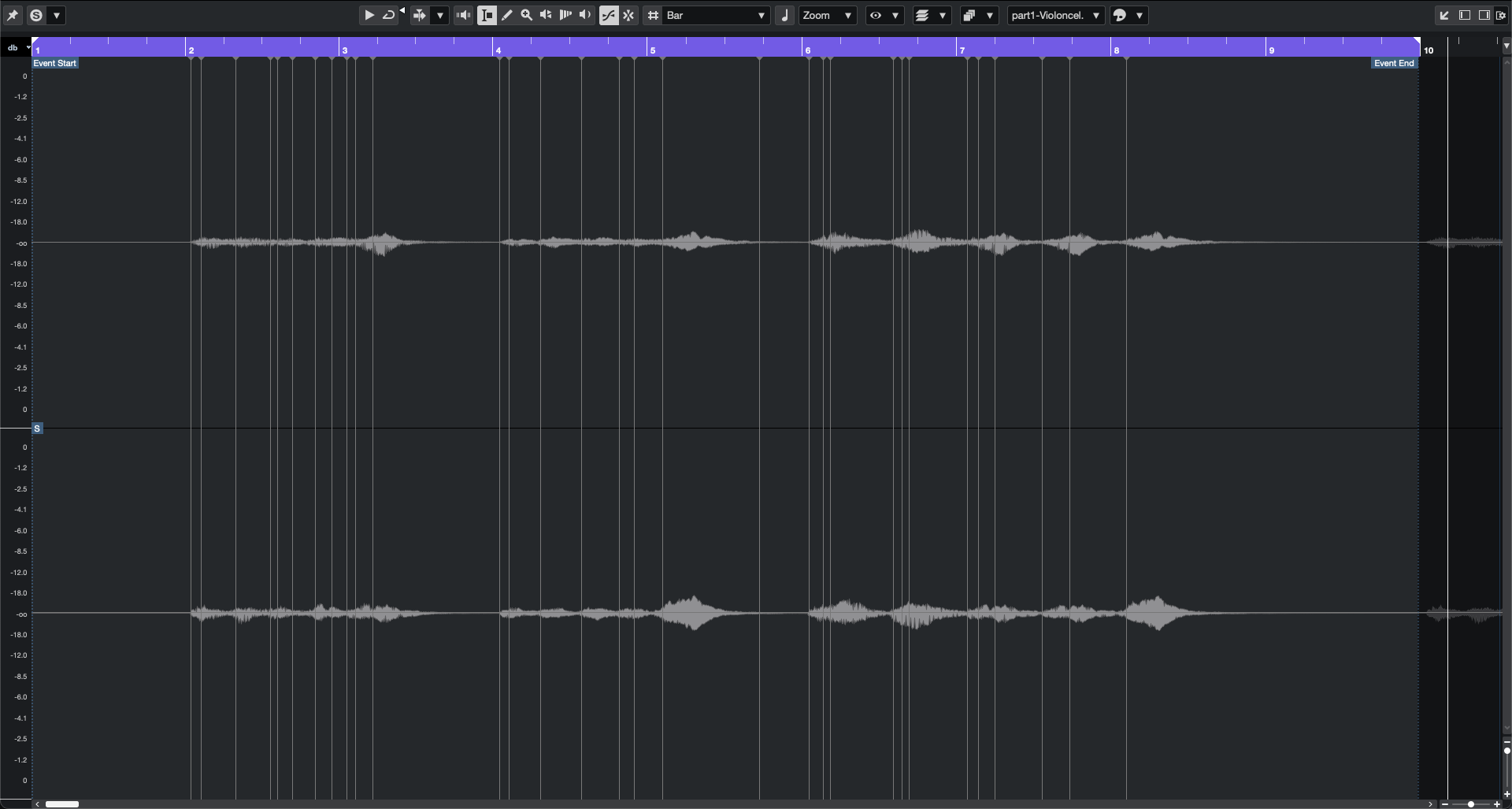}}
    \caption{Opposing states of transient (Hitpoints) trigger density between the dominant melody and the bass foundation.}
    \label{fig:hitpoints_main}
\end{figure}

\subsection{Falsification of "Mechanical Determinism" via Loudness Features}
In traditional AI feature extraction engineering, there is often an oversimplified "mechanical assumption": the voice with the highest loudness (highest RMS) must be the main melody. However, conducting a complete AES-17 RMS and EBU R128 loudness measurement \cite{ref10} on the first 8 measures of \textit{The Lark} (see Table \ref{tab:rms_data}) yields counter-intuitive results: the cello, acting as the accompaniment foundation, has a significantly higher average RMS (-34.6 dB) than the first violin, which is the absolute core (-40.1 dB).

\begin{table}[t]
\centering
\caption{Loudness and Dynamic Features of the Four Voices in the Opening of \textit{The Lark}}
\label{tab:rms_data}
\resizebox{\linewidth}{!}{%
\begin{tabular}{llccc}
\toprule
Role & Voice & RMS(dB) & LUFS & Loudness Range(LU) \\
\midrule
Leading Core & First Violin & -40.1 & -33.1 & 6.0 \\
Groove Base & Cello & -34.6 & -33.6 & 4.6 \\
Harmonic Filler & Second Violin & -40.9 & -40.4 & 1.3 \\
Harmonic Filler & Viola & -42.3 & -41.3 & 6.7 \\
\bottomrule
\end{tabular}%
}
\end{table}

This contrast originates from the naturally higher energy density possessed by low-frequency sound waves. It directly falsifies the mechanical determinism that "high loudness equals leading voice" at a physical level. It profoundly reveals that in classical multi-part interactions, establishing a core role does not rely on simple physical loudness suppression, but on a comprehensive contrast of frequency bands and transients. Furthermore, the extremely low integrated loudness (-40.4 LUFS) and loudness range (1.3 LU) of the second violin perfectly corroborate its restrained "harmonic filler" attribute. This data-driven conclusion demonstrates that when reconstructing representation spaces for machine learning, we must resolutely move towards a "phenomenological paradigm" of multi-dimensional acoustic feature fusion.

\section{Cross-Modal Representation: Low-Level Reconstruction as "Phenomenological Anchors"}
Pure symbolic music data (such as Note On / Note Off in the MIDI protocol) \cite{ref11} contains only discrete mathematical dimensions like pitch and duration. This data structure represents the purely rational "symbolic brain" of Cartesian mind-body dualism, abstracting away all physical weight of music as acoustic vibration. This study attempts to inject multi-dimensional information into single information symbols through the intervention of electroacoustic data, achieving cross-modal information neutralization.

\subsection{Rejecting Mechanical Determinism: Continuous Feature Space as Aesthetic Heuristic}
In highly complex classical textures, the roles of voices are fluid, polysemous, and full of metaphors. Based on the empirical loudness data presented above, relying on fixed frequency response thresholds to categorically classify musical roles will inevitably fall into the theoretical trap of mechanical determinism.

Therefore, this study completely abandons absolute determination logic during the preprocessing stage. Instead, continuous energy envelopes and transient trigger densities extracted earlier undergo dimensionality reduction and feature normalization \cite{ref11}. These acoustic quantities, originally existing in a continuous physical space, are transformed into soft information—"Phenomenological Anchors" and "Aesthetic Heuristics"—and fused with discrete symbolic logic data. At this point, the machine is no longer simply fitting a pitch number; it establishes a numerical perception of the explosive force and expansiveness of that note in physical space.

\subsection{Feature Space Reconstruction and System Alignment Based on Event Timestamps}
The fixed time grid (e.g., sixteenth-note quantization) adopted by traditional models often erases the micro-timing present in real performances. Consequently, this paper abandons this architecture in favor of Event-based Timestamps. By extracting accompaniment transients, the system introduces a relative time offset system ($\Delta t$) \cite{ref4}. To prevent cumulative drift during multi-part interactions in the DAW, $\Delta t$ discards independent absolute timers, strictly binding to the host pulse resolution (PPQ, Pulses Per Quarter Note) for relative difference synchronization.

Guiding the model to directly learn PPQ-based event trigger differences is structurally more aligned with the underlying characteristics of deep temporal networks, circumventing cumulative errors caused by forced gridding. Aesthetically, it liberates musical time from homogenous, mechanical "clock time," restoring it to the "duration" (Durée) defined by Henri Bergson. This dual technical and philosophical reconstruction ensures that the true micro-dynamics of voice interaction are precisely preserved in the digital dimension with a "phrase breathing" feel.

\begin{figure}[t]
    \centering
    \includegraphics[width=\linewidth]{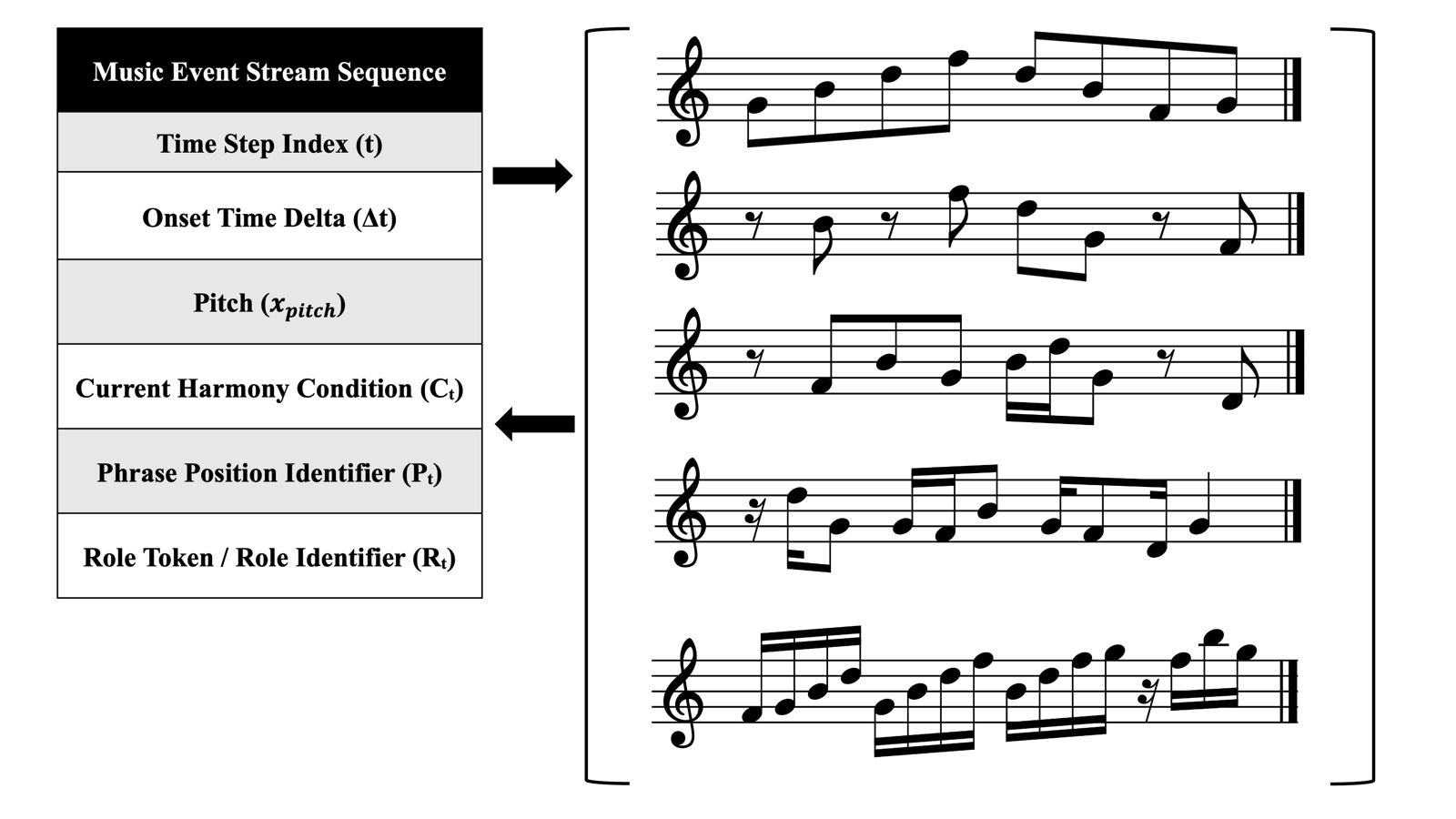}
    \caption{Logical architecture diagram of multi-dimensional machine representation based on role-aware encoding and an elastic timeline.}
    \label{fig:role_architecture}
\end{figure}

To clearly articulate the multi-dimensional input vector space constructed in this study, Fig. \ref{fig:role_architecture} illustrates the underlying feature extraction logic. As the architecture diagram shows, within the low-level structure of the musical event sequence, aside from preserving traditional pitch, phrase position, and the Current Harmony Condition, the system forcefully embeds the aforementioned "PPQ-based host-synchronized time offset ($\Delta t$)" and the "Role-Aware Encoding ($R_t$)." The core purpose of this composite feature space reconstruction is to enable deep temporal neural networks to perceive fluid role contexts and micro-elastic temporal features in real time when fitting polyphonic generation probabilities.

\begin{table*}[t]
\centering
\renewcommand{\arraystretch}{1.5}
\setlength{\tabcolsep}{18pt}
\caption{Machine Representation Heuristic Lookup Table Based on Acoustic Feature Mapping}
\label{tab:role_mapping}
\begin{tabularx}{\textwidth}{l X X X}
\toprule
\textbf{Defined Role} & \textbf{Phenomenological Anchor (Comprehensive Physical Profile)} & \textbf{Morphological Description (Classical Auditory Qualitative)} & \textbf{High-Dimensional Rep. (Role Embedding)} \\
\midrule
Leading Core & High-frequency prominence, extremely low transient density, smooth long ADSR envelope & High register, continuous and singing melodic line & $E_{\text{role}} = [1, 0, 0, 0]$ \\
Groove Base & Ultra-low frequency standing waves, extremely high transient density, high average RMS & Predominantly staccato, high pitch density, maintains beat & $E_{\text{role}} = [0, 1, 0, 0]$ \\
Harmonic Filler & Mellow mid-frequency standing waves, low loudness fluctuation (small Loudness Range) & Sustained long notes, fills the mid-frequency acoustic space & $E_{\text{role}} = [0, 0, 1, 0]$ \\
Contrapuntal Response & Transient triggers alternate and mutate as the leading voice decays & Fills motifs during leading voice rests to form imitation & $E_{\text{role}} = [0, 0, 0, 1]$ \\
\bottomrule
\end{tabularx}
\end{table*}

\subsection{Mathematical Representation and Real-Time Inference of Role-Aware Encoding}
Integrating the physical texture anchors listed in Table \ref{tab:role_mapping}, this study assigns an additional "role-aware encoding" dimension $E_{\text{role}}$ to every independent note event via unsupervised cross-modal annotation. When constructing the input nodes of the deep temporal neural network, the composite feature vector $x_t$ of a specific voice at the current time step $t$ is parametrically expressed aesthetically as:

\begin{equation}
\begin{aligned}
x_t = \text{Concat}\big(E_{\text{pitch}}(t), E_{\text{vel}}(t), E_{\text{role}}(t), \Delta t\big)
\end{aligned}
\end{equation}

The mandatory injection of the $E_{\text{role}}$ dimension endows the system with an a priori clue to understanding the social attributes of polyphonic music, forcing the model to internalize the yielding logic between role state machines during backpropagation. Furthermore, to support the ultra-low latency interaction of the aforementioned high-dimensional feature tensors in real-time environments, the system's inference backend abandons high-overhead frameworks like ONNX. Instead, it deploys the RTNeural network library specifically designed for audio DSP \cite{ref12}, enabling the model to make responses with an awareness of compromise and yielding within millisecond-level latencies.

\section{Interactive Aesthetics of Human-Computer Collaboration: The Philosophical Spirit and Contemporary Translation of \textit{The Lark} Quartet}
The artistic value of Haydn's \textit{The Lark} String Quartet goes beyond textural techniques and melodic expression; it lies more in the Enlightenment philosophical logic embedded within its four-voice structure—the spirit of an egalitarian community, the dialectic of freedom and order, and the enduring rhythms of natural life together constitute the work's spiritual core. As a paradigm of Haydn's mature string quartets, this work has long been regarded as a representative text of the high unity of classical chamber music form and spirit \cite{ref13}. Reactivating this classical logic from a contemporary technological perspective reveals that the interactive paradigms accumulated over centuries in chamber music provide a solid philosophical anchor for contemplating the musical aesthetics of human-computer collaboration. This section reflects on the deep philosophical connotations of human-machine collaborative music from three dimensions: the ethics of the subject, the ontology of time, and the artistic paradigm.

\subsection{Ethics of Interacting Subjects: The Voice Community and the Aesthetic Foundation of "Otherness Awareness"}
Music is essentially a social auditory practice, a fundamental consensus in musical aesthetics \cite{ref14}; the string quartet is the most exquisite artistic vehicle for this social nature, considered the highest manifestation of the dialogic spirit in instrumental music \cite{ref15}. In the classical quartet paradigm established by \textit{The Lark}, the four instruments break free from the "homophony + accompaniment" hierarchical relationship of the Baroque era, forming a community of equal, interdependent voices. This is the musical manifestation of the Enlightenment idea of the "dialectical unity of freedom and order." The free singing of the first violin in the exposition is always bounded by the stable order of the lower three voices; in the development, the leadership dynamically circulates among the four voices, with no eternal center or absolute periphery. The free expression of each voice is always predicated on listening to and yielding to others. This tacit understanding is essentially an "interactive rationality" and "voice ethics" in music.

This ethic is precisely the core lacking in traditional AI music generation. In existing symbolic generation logic, the model is always a "monologue subject," outputting notes unidirectionally based solely on prior probability matrices, completely devoid of perception and response to "other voices." This fundamentally contradicts the interactive essence of polyphonic music as a multi-subject dialogic symbiosis. When a machine solely focuses on its own melodic output and ignores the state of other voices in the texture, the polyphonic text it generates possesses form but lacks the soul of a dialogue.

Therefore, adding a role-perception dimension to the low-level representation of music generation holds far greater philosophical value than mere engineering optimization. Its core significance lies in allowing digital generation systems to transition, for the first time, from "being-in-itself" to "being-for-others," acquiring a fundamental "otherness awareness." The system no longer outputs blindly based on local pitch probabilities but can identify the role division of the overall texture and clarify its functional positioning. The establishment of this role cognition is the reconstruction of classical chamber music voice ethics in digital space: freedom is not boundless catharsis, expression is predicated on listening, and the aesthetic value of the individual can only be truly realized in the harmony of the community.

The ultimate goal of human-computer collaboration is to allow the machine to truly enter the interactive community of music, becoming a dialogic partner with basic ethical awareness. Only when the machine understands yielding, receiving, and role handover can human-machine ensemble playing possess the most precious "sense of dialogue" found in chamber music.

\subsection{Ontology of Temporality: Disenchantment of the Mechanical Grid and the Return of Musical Duration}
Time is the mode of existence of music, the fundamental dimension upon which musical art unfolds \cite{ref14}; understanding time directly determines the spiritual undertone of musical art. Dominated by industrial rationality and technical thinking, musical time has long been reduced to precisely quantifiable physical clock time. The fixed note grid widely adopted in traditional symbolic music generation is exactly the product of this mechanistic view of time: time is abstracted into homogeneous, infinitely divisible mathematical units, and every note is pinned to an absolutely precise scale. This approach prioritizes engineering efficiency, but it dissolves the life-like texture of musical time, alienating vivid artistic experience into regulated industrial products.

This mechanistic view of time runs contrary to the natural life philosophy carried by \textit{The Lark}. The time in \textit{The Lark} is the life time of natural organisms: the singing has a natural breath, the melody has internal tension and relaxation, the phrasing has subtle breathing room; it follows the growth rhythm of natural things and the natural flow of the performers' bodily perceptions. The philosophy of "returning to nature" implicit in the work signifies that true musical order is never externally forced regularity, but the natural duration of inner life, deeply echoing Rousseau's Enlightenment aesthetic propositions \cite{ref16}.

From an ontological perspective, abandoning the fixed grid and adopting an elastic time representation based on event triggers is essentially a return and redemption of musical temporality. It no longer disciplines musical time with abstract mathematical scales, but defines time flow by the sequential relationships and interval tensions of sound events. This echoes profoundly with Henri Bergson's concept of "duration" (Durée): musical time is not spatializable, fragmentable physical time, but a continuous continuum of conscious experience, a qualitative duration containing strength, urgency, and tension \cite{ref17}. Event-based temporal representation precisely restores the true face of duration in the digital dimension, shifting the machine's perception of time from "calculated time" to "experienced time."

This reconstruction of temporality also responds to the embodied nature of musical performance. Its philosophical foundation traces back to Merleau-Ponty's phenomenology of perception—the body is not a cognitive object, but the primal field where cognition occurs \cite{ref18}. Empirical research on musical performance also confirms that micro-temporal processing, such as tempo rubato, is the result of the coordinated externalization of the performer's body movements and emotional experiences, a direct product of embodied cognition \cite{ref19}. For a machine without bodily experience, learning event time differences is essentially imitating and internalizing human life rhythms, giving the generated music human "warmth." It is also a contemporary echo of classical music's humanistic spirit.

\subsection{Paradigm Shift: From Instrumental Rationality to an Aesthetic Community of Human-Machine Symbiosis}
The role perception dimension constructs the spatial ethics of human-computer interaction, and the elastic time representation reconstructs the temporal foundation of musical existence. Together, these two drive artificial intelligence music to complete a critical philosophical paradigm shift: from "instrumental music production" to "symbiotic interactive composition." This leap is a micro-projection of the broader trend of generative AI deeply intervening in music composition within the realm of classical chamber music. As revealed by the concept of "Machinism" proposed by Xiaobing Li: when generative capabilities transition from external tools to structural embeddings of the composition system, the subject morphology and meaning production mechanism of music creation are undergoing systemic reconstruction. This theory provides an exact academic anchor for understanding the aesthetics of human-computer collaborative chamber music \cite{ref20}.

The core of "Machinism" does not advocate for machines replacing humans as independent creative subjects; rather, it posits that under human value settings and aesthetic constraints, intelligent systems can be viewed as structural collaborative forces in meaning production, becoming an extension of human emotional expression and aesthetic organization, forming an "extended subjectivity": human aesthetic judgment and value orientation remain the source of meaning, while intelligent systems participate in meaning generation with their structural computing capabilities, becoming organizers and amplifiers of expression. This study's reconstruction of the low-level representation of \textit{The Lark} quartet is exactly a concrete practice of Machinism in the dimension of classical music composition: transforming the voice interaction logic, role flow order, and elastic time rhythm precipitated in the work into structural parameters that the system can recognize and follow. Humans remain the setters of aesthetic goals and the dominators of interactive ethics, while the machine uses digital computing power to extend the interactive aesthetics of classical chamber music into digital space.

The construction of this extended subjectivity forms a cross-temporal echo with the Enlightenment community spirit carried by \textit{The Lark}. Charles Rosen points out in \textit{The Classical Style} that Haydn's establishment of the string quartet paradigm was itself a "decentralized" expansion of the musical subject, bringing four instruments into an equal dialogic community \cite{ref6}. Paul Henry Lang in \textit{Music in Western Civilization} also suggests that Haydn's quartet innovation was an acoustic projection of the Enlightenment's social ideals of equality and rationality, and the balanced four-voice structure is an acoustic microcosm of the zeitgeist \cite{ref21}. The extended subjectivity advocated by Machinism is yet another boundary expansion in the digital age: breaking the inherent notion that "the creative subject can only be a biological individual," bringing intelligent systems into the music composition community. The internal logic of the two is highly consistent: both reject absolute subject hegemony and believe in the rational spirit of equal collaboration and harmony in diversity.

From the perspective of Machinism, this paradigm shift is a historical advancement of "extended humanism." As Xiaobing Li states, the core value of machine participation in creation lies not in the machine possessing independent emotions, but in humans making technology an extension and amplifier of emotional communication by setting aesthetic goals and defining value scales \cite{ref20}. The ultimate goal of reconstructing machine representation is to allow the interactive wisdom and aesthetic spirit of classical chamber music to break through professional technical barriers, gaining broader dissemination space, and allowing more people to experience the beauty of ensemble playing. Its essence is the popularization and extension of the humanistic spirit.

From the pastoral dialogue of four violins to the cross-subject collaboration between humans and digital musicians, the core spirit of music remains unchanged—it is always an art of listening, dialogue, and community. The lark singing freely within order in \textit{The Lark} is not only the artistic incarnation of the Enlightenment humanistic ideal but also illuminates the ultimate direction of human-machine collaborative aesthetics: guarding the agility of life within a rational framework, steadfastly preserving the humanistic undertone in a technological context, and ultimately arriving at a harmonious symbiosis between individual and whole, human and machine, technology and art.

\section{The Aesthetic Paradigm of Human-Computer Collaboration: Technical Translation and Humanistic Boundaries}
Using \textit{The Lark} as a paradigm, the preceding text argued the aesthetic core of role-aware encoding and elastic time representation from three levels: voice ethics, temporal ontology, and paradigm shift. This research is not merely algorithmic optimization, but a digital translation of the interactive spirit of classical chamber music accomplished using technology as a carrier. When the technical path of instrumental rationality returns to the aesthetic anchor of the musical ontology, human-computer collaboration is no longer just an improvement in production efficiency; it points to a completely new musical aesthetic paradigm. With the mutual construction of philosophy and technology as the core thread, this section offers an aesthetic-oriented prospect of the extension direction of this paradigm, reflects on its inherent boundaries and limitations, and ultimately returns to the core stance of "technology serving humanities."

\subsection{The Aesthetic Return of Technology: Representation Reconstruction Oriented towards Musical Ontology}
Throughout the development of AI music, technology has long existed with an attitude of instrumental rationality: fixed-grid time representations and single-subject generative logic both aimed at production efficiency as their core goal, inadvertently deviating from the ontological attributes of music. The core value of the low-level representation system built in this paper lies exactly in correcting this deviation—it uses structural technological language to recall the essential definition of music as an "art of dialogue" and "art of time."

In the dimension of the subject, the technical design of role-aware encoding is fundamentally the digital implementation of classical "voice ethics." It transforms the interactive rules of "listening to others, yielding and receiving, and role flow" in chamber music into structural rules that the system can recognize and execute. It grants basic otherness awareness to one-dimensional generative logic, implants the intersubjective premise of polyphonic music into the low-level framework of symbolic generation, and shifts technology from a "monological production tool" to a "dialogical interactive carrier."

In the temporal dimension, elastic time representation based on events is essentially a structural restoration of "musical duration" \cite{ref17}. It abandons the homogenous time view of mechanism, defines time flow by the relational nature of sound events, and returns musical time from quantifiable physical scales to a qualitative experience of duration. This is both an echo of Bergson's life philosophy in the field of digital music and a technical response to the embodied nature of musical performance \cite{ref18, ref19}—it does not pursue absolute precision of the beat but attempts to restore the life texture of musical flow, bringing the machine's perception of time closer to human life experience.

In short, the technical path of this paper is to carry classical aesthetic cores via digital structures, allowing technology to no longer remain external to art but to become a way for musical ontology to extend into digital space.

\subsection{The Aesthetic Extension of the Paradigm: The Creative Landscape of Human-Machine Symbiosis}
When technology anchors its aesthetic foundation, human-computer collaboration elevates from scattered technical attempts to a new musical composition paradigm with universal significance. The core of this paradigm is the "extended subjectivity" revealed by "Machinism": humans are always the source of aesthetic value, while intelligent systems act as structural collaborative forces participating in meaning production \cite{ref20}. The aesthetic extension direction is primarily reflected on three levels.

First, the reconstruction of the creative subject and the shift of aesthetic focus. As generative capabilities become embedded in composition systems, the core value of creation will shift from "the ability of manual writing" to "the ability of aesthetic judgment and system organization." Creators will not need to write every voice note by note, but will focus more energy on setting aesthetic goals, designing interaction logic, and controlling overall texture. This is not the dissolution of the creator's subjectivity, but its upgrade—humans are liberated from operational labor and return to the core aesthetic decision-making level of artistic creation.

Second, the popularization and vitalization of classical interactive wisdom. The dialogic spirit of chamber music has long been restricted by professional skill thresholds, achievable only among a minority of professional players. Structured role interaction logic can transform hidden ensemble rules into callable, deconstructible aesthetic frameworks, lowering the entry barrier for the art of voice dialogue. Whether in ensemble teaching, classical interpretation, or contemporary composition, the interactive wisdom accumulated in classical chamber music will use digital media to break through professional boundaries and penetrate into broader musical practice scenarios.

Third, the expansion of the boundaries of the musical community. From the equal dialogue of four voices in the Enlightenment era to the collaborative symbiosis between humans and intelligent systems in the digital age, the form of the musical community constantly expands, but the core spirit of "equal dialogue, harmony in diversity" remains unchanged. Human-computer collaboration is not using technology to replace humans, but introducing new participatory dimensions to musical dialogue. It extends the musical community from human-to-human social connections to a composite subject structure of human-machine collaboration, continuing the social essence of music in a new era context.

\subsection{Boundaries and Reflections: The Aesthetic Limits of Technological Intervention}
First is the limit of stylistic paradigm adaptation. The interaction logic and representation system of this paper are abstracted from Classical-era string quartets with clear tonality and explicit textural layering, presupposing the classical rational view of order and role-differentiated textural logic. It cannot directly cover all musical forms: the highly fused symphonic textures of late Romanticism, the sound experiments of the 20th-century avant-garde that dissolved primary-secondary relationships, or free improvisation practices without preconditions, are all difficult to regulate with this structured role framework. Elevating the aesthetic logic of a specific style to a universal standard would only make technology a new formal dogma, narrowing the diverse possibilities of musical art.

Second is the essential difference at the ontological level. Technology can simulate the form of role interaction and fit the morphology of time duration, but it cannot possess true embodied life experience \cite{ref18}. Role-aware encoding achieves structural cognitive otherness, not empathetic interaction based on bodily experience and emotional resonance; elastic time representation restores rhythmic elasticity in a statistical sense, not the unique temporal texture originating from individual life experience \cite{ref17}. The depth and temperature of music are ultimately rooted in human life experience and the spiritual world; this is the ontological core that statistical fitting of algorithms can never reach.

Finally, there is the irreplaceable nature of humanistic value. The community spirit of music is essentially human-to-human social connection and spiritual resonance, which is the core humanistic value of chamber music art. Human-computer collaboration can build structural frameworks for voice interaction but cannot replace the emotional tacit understanding and spiritual connection achieved between humans in an ensemble. Intelligent systems can become efficient compositional aids and practice tools, but they can never replace the deep emotional connections and social identification established through music between people in face-to-face ensemble playing.

In conclusion, this paper's exploration of human-computer collaborative aesthetics is essentially a digital translation of the spirit of classical music: using technology as the path, aesthetics as the anchor, and humanities as the destination. Technology is always an extended carrier of the humanistic spirit, not a replacement for art. Only by recognizing the boundaries of technology and maintaining human aesthetic subjectivity can the dialogic spirit of music truly continue in the digital age, continuously generating new artistic possibilities in the context of human-machine symbiosis.

\section{Conclusion}
Starting with a specific interdisciplinary analysis of Haydn's string quartet \textit{The Lark}, this paper completes an exploratory journey from auditory qualitative analysis and electroacoustic mapping to the low-level representation reconstruction in machine learning. By reducing the dimensionality of physical features captured through electroacoustic quantification and transforming them into Event-based Timestamps and Role-Aware Encoding, this study methodologically breaks the "mechanical determinism" and "perception blind spots" common in pure symbolic music generation. This cross-modal feature extraction logic not only endows rigid digital matrices with the elasticity of temporal duration but also injects a priori clues for understanding voice interaction and power yielding into the machine learning system at the low-level data structure. As embodied intelligence and low-latency interaction architectures continue to advance, AI music built on multi-dimensional feature perception is bound to evolve into collaborative creative subjects with deep conversational and social listening capabilities, opening a new era of human-machine co-creation.

This study also has certain limitations: at the current stage, it only uses a single quartet excerpt from Haydn's \textit{The Lark} as the analysis sample and has not yet covered the more complex polyphonic textures of late Romantic or modern music. Moreover, an end-to-end training of the deep model to verify the actual generation performance of this encoding has yet to be completed. Future research will further expand the datasets to include classical and modern music of multiple genres and conduct full engineering experimental validations.

\FloatBarrier

\appendices

\section{Complete Supplement to Experimental Mapping of Physical Fingerprints}
To support the "phenomenological anchors" discussed in Section III, this appendix presents a complete independent electroacoustic mapping profile of the four voices in measures 1-8 of the first movement of Haydn's \textit{The Lark}, verifying the objective rigor of physical role isolation within the human-computer collaborative aesthetic framework.

\subsection{Independent Spectral Observation Data of Each Voice}
\begin{figure}[t]
    \centering
    \subfloat[First Violin Independent Spectrum]{\includegraphics[width=0.48\linewidth]{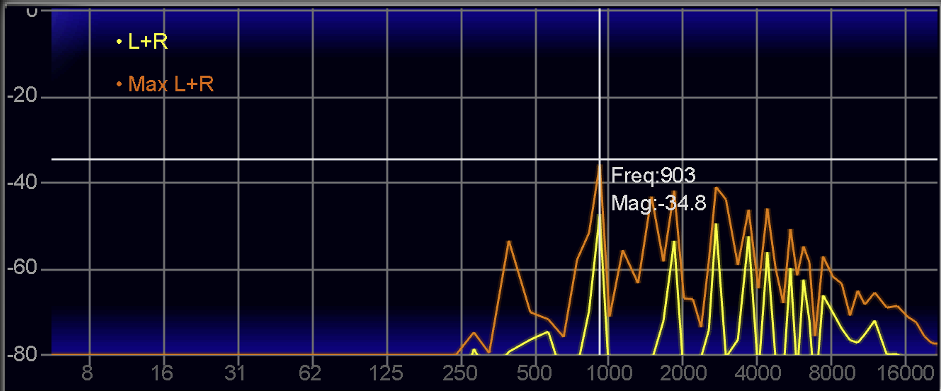}}
    \hfill
    \subfloat[Second Violin Independent Spectrum]{\includegraphics[width=0.48\linewidth]{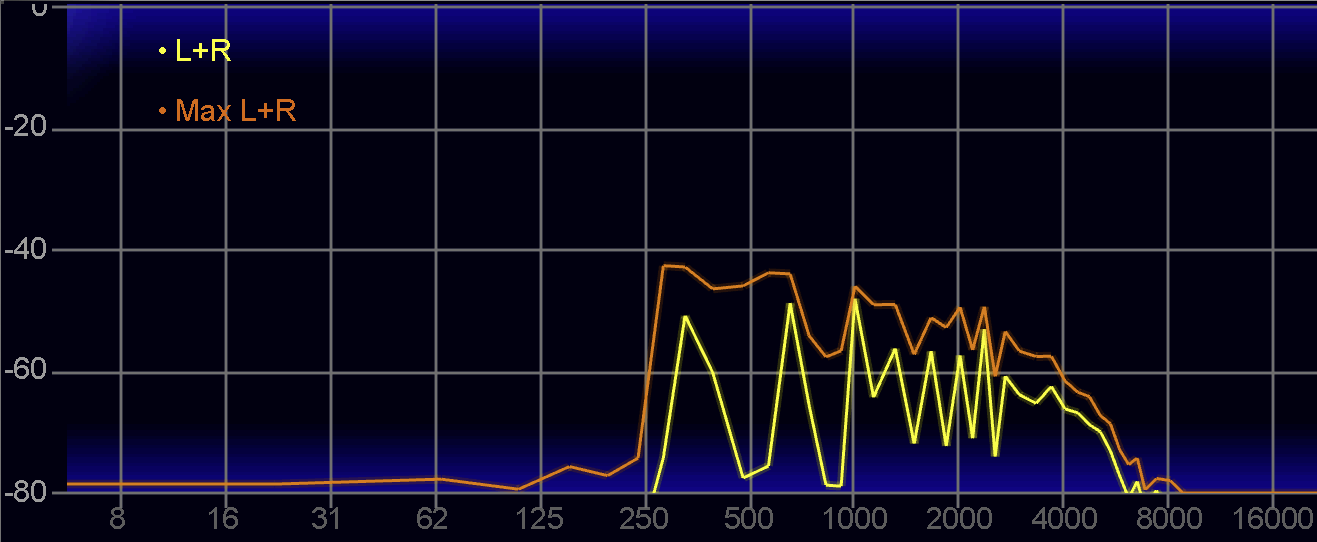}}
    
    \vspace{0.2cm}
    
    \subfloat[Viola Independent Spectrum]{\includegraphics[width=0.48\linewidth]{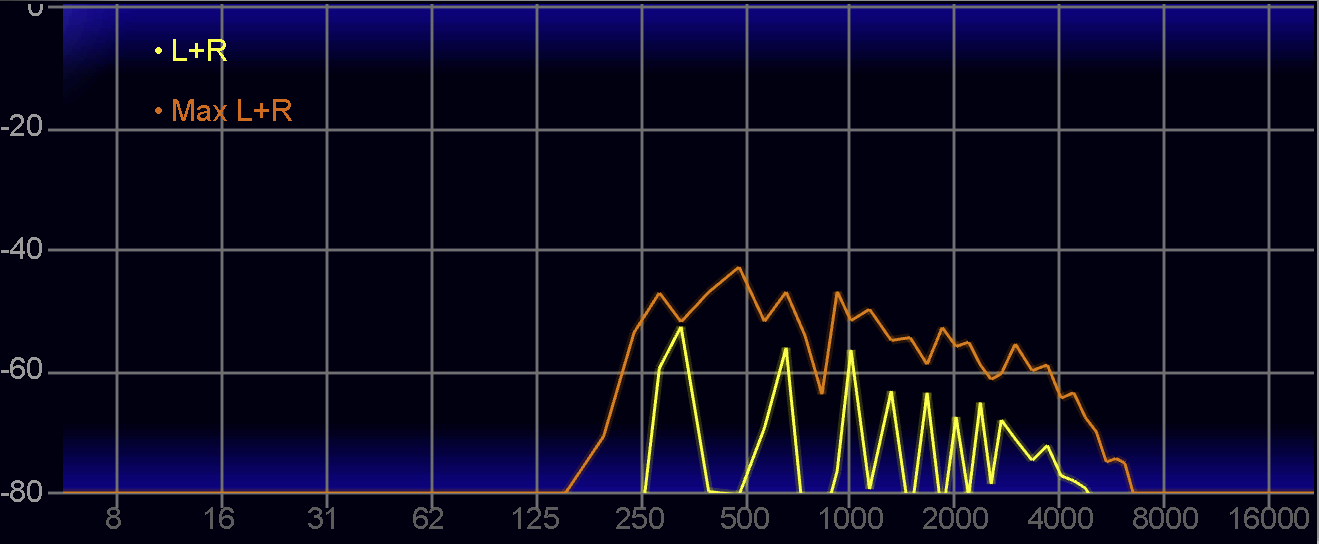}}
    \hfill
    \subfloat[Cello Independent Spectrum]{\includegraphics[width=0.48\linewidth]{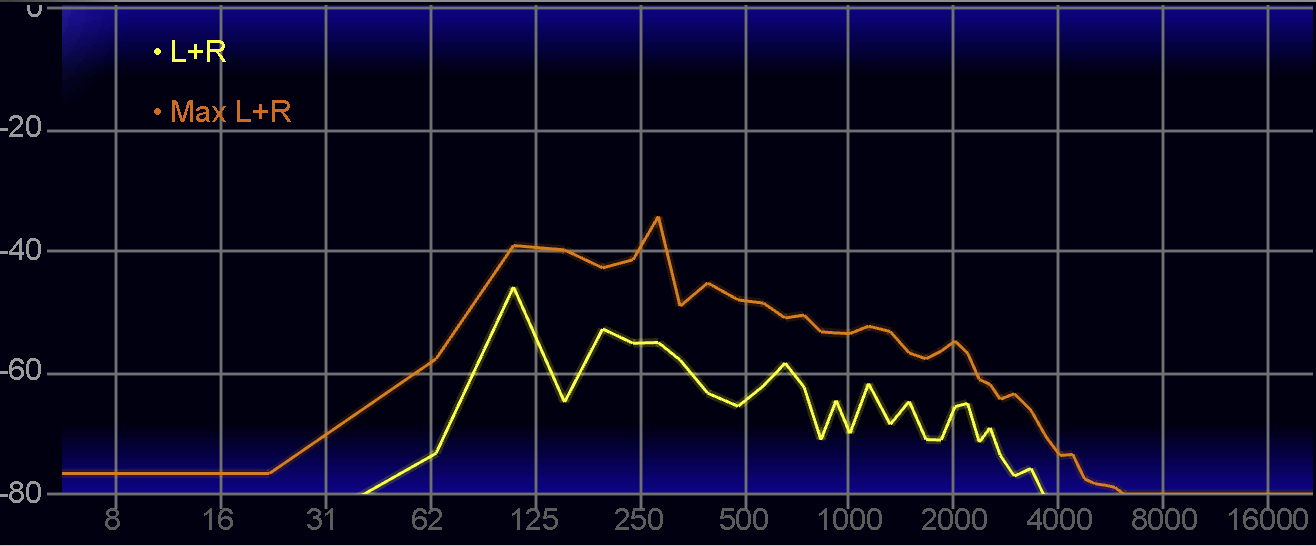}}
    \caption{Supplementary profiles of the independent frequency response envelopes of the four string voices.}
    \label{fig:appendix_spectrum}
\end{figure}

As shown in Fig. \ref{fig:appendix_spectrum}, the independent spectral mapping of each track clearly reveals a stepwise descent of the energy center of gravity from high to low. The mid-frequency standing waves of the viola and second violin form a stable internal harmonic structure.

\subsection{Complete Observation of Time-Domain Transients (Hitpoints) and ADSR Envelope Comparison}
The inner voices (second violin and viola) exhibit high-density transient fluctuation characteristics similar to the cello at this stage, further supporting the feasibility of using time-domain trigger frequency as a criterion for the "accompaniment foundation." Meanwhile, the typical single-note time-domain envelope (ADSR) presents the most direct physical mapping of the auditory "sweeping" and "short" characteristics.

\begin{figure}[t]
    \centering
    \subfloat[Second Violin Transients (High-frequency spiccato)]{\includegraphics[width=\linewidth]{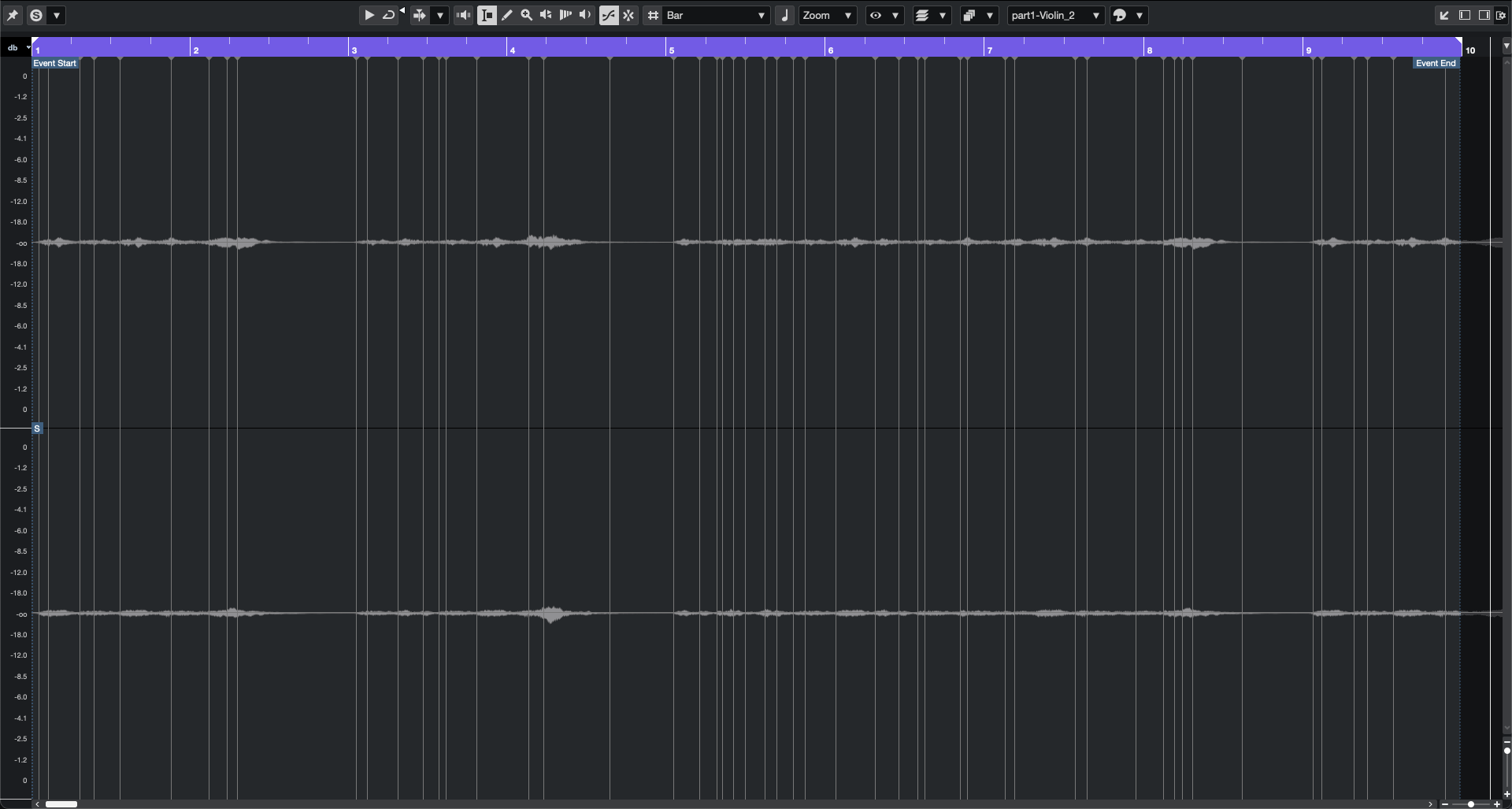}} \\
    \vspace{0.2cm}
    \subfloat[Viola Transients (Mid-frequency arpeggiation)]{\includegraphics[width=\linewidth]{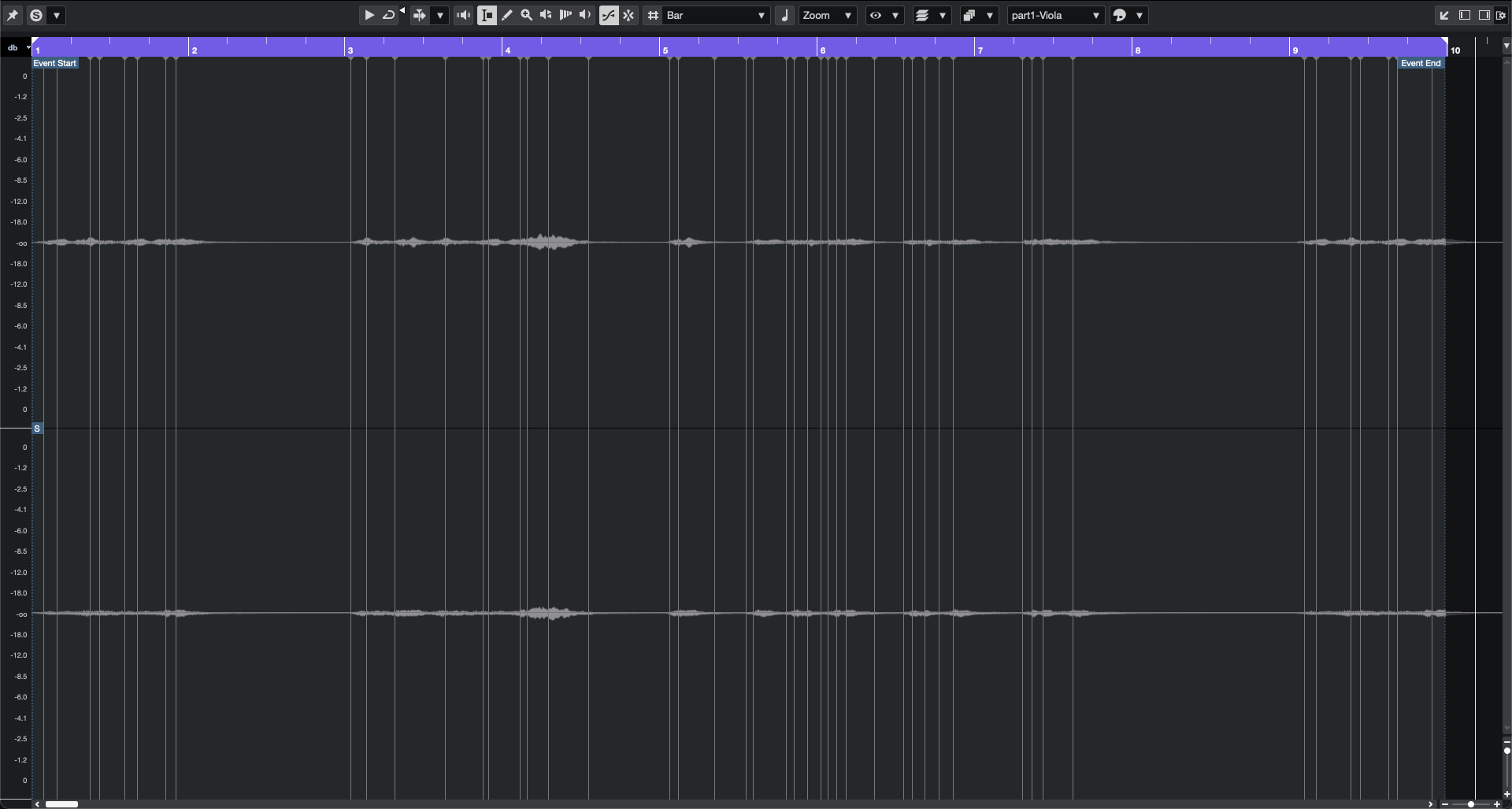}}
    \caption{Transient (Hitpoints) mapping profiles of the inner harmonic voices.}
    \label{fig:appendix_hitpoints}
\end{figure}

\begin{figure}[t]
    \centering
    \subfloat[Leading Melody: Extremely slow, long sustain envelope]{\includegraphics[width=\linewidth]{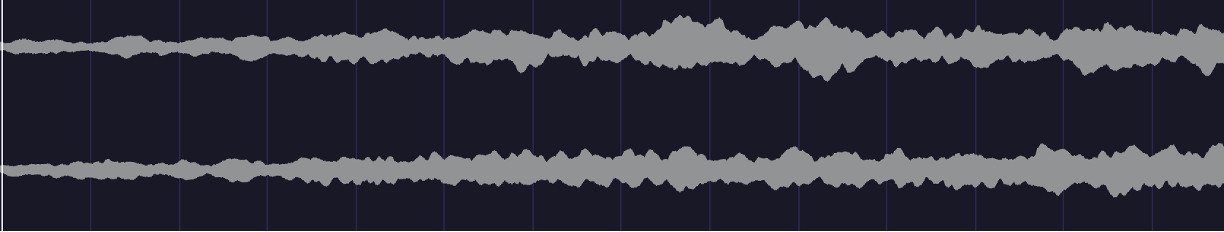}} \\
    \vspace{0.2cm}
    \subfloat[Groove Base: Instant attack, fast decay envelope]{\includegraphics[width=\linewidth]{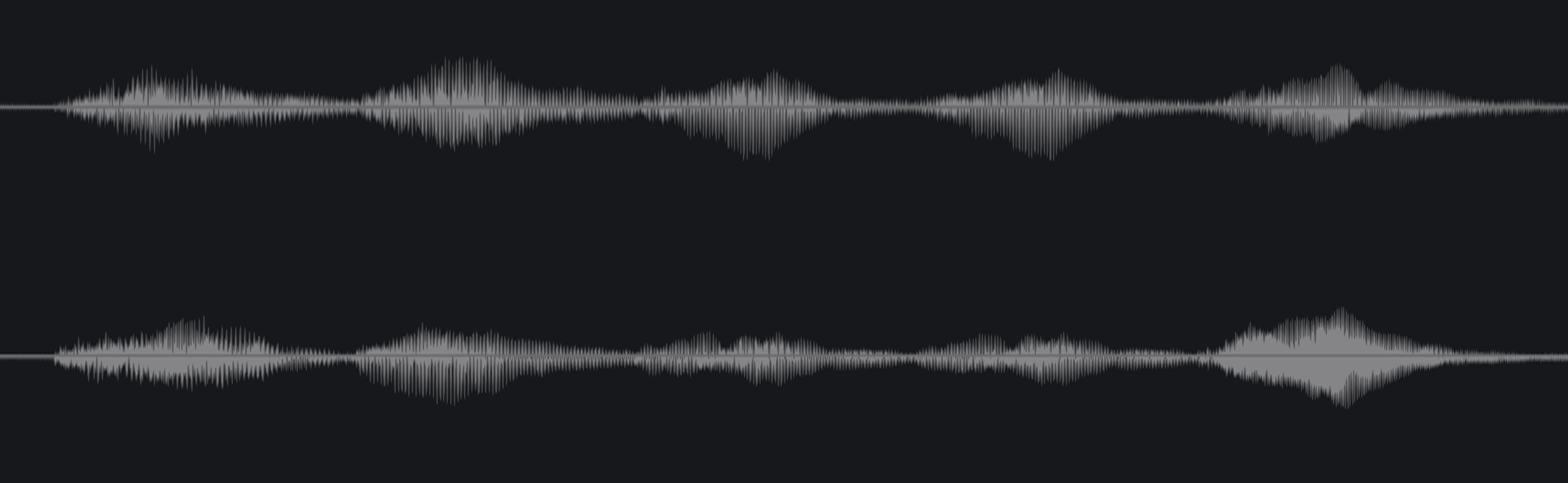}}
    \caption{Comparison of typical single-note time-domain envelopes (ADSR) of core opposing roles.}
    \label{fig:appendix_adsr}
\end{figure}



\begin{thebibliography}{99}
\bibitem{ref1} A. D'Ausilio, et al. "Measuring social interaction in music ensembles," Philosophical Transactions of the Royal Society B, 370(1664), 2015.
\bibitem{ref2} H.-W. Dong, et al. "MuseGAN: Multi-track sequential generative adversarial networks for symbolic music generation," in AAAI, 2018.
\bibitem{ref3} C.-Z. A. Huang, et al. "Music Transformer," in ICLR, 2019.
\bibitem{ref4} S. Oore, et al. "This time with feeling: Learning expressive musical performance," Neural Computing and Applications, 32(4), 2020.
\bibitem{ref5} G. Widmer, "Machine discoveries: A few simple, robust local expression principles," Journal of New Music Research, 31(1), 2002.
\bibitem{ref6} C. Rosen, The Classical Style: Haydn, Mozart, Beethoven. W. W. Norton \& Company, 1997.
\bibitem{ref7} W. D. Sutcliffe. The String Quartets of Joseph Haydn. Cambridge University Press, 2002.
\bibitem{ref23} W. E. Caplin. Classical Form: A Theory of Formal Functions for the Instrumental Music of Haydn, Mozart, and Beethoven. Oxford University Press, 1998.
\bibitem{ref24} P. Griffiths. The String Quartet: A History. Thames and Hudson, 1983.
\bibitem{ref25} R. Qian, "Haydn's 'Lark' Quartet," Music Lover, no. 3, 1982.
\bibitem{ref8} M. Müller. Fundamentals of Music Processing. Springer, 2015.
\bibitem{ref9} J. P. Bello, et al. "A tutorial on onset detection in music signals," IEEE Transactions on Speech and Audio Processing, 13(5), 2005.
\bibitem{ref10} European Broadcasting Union. "EBU R 128: Loudness normalisation and permitted maximum level of audio signals," 2020.
\bibitem{ref11} A. Ycart and M. Benetos. "A study on LSTM networks for polyphonic music sequence modelling," in ISMIR, 2017.
\bibitem{ref12} J. Chowdhury. "Real-time neural network inferencing for audio DSP," in DAFx, 2021.
\bibitem{ref13} F. K. Grave and M. G. Grave. The String Quartets of Joseph Haydn. Oxford University Press, 2006.
\bibitem{ref14} R. Yu. General Theory of Music Aesthetics. Shanghai Music Publishing House, 2000.
\bibitem{ref15} Y. Yang. "The Spiritual Implications and Cultural Value of Chamber Music," People's Music, no. 8, 2017.
\bibitem{ref16} J.-J. Rousseau. Discourse on the Sciences and Arts. (Trans.) Commercial Press, 1963.
\bibitem{ref17} H. Bergson. Time and Free Will: An Essay on the Immediate Data of Consciousness. (Trans.) Commercial Press, 1958.
\bibitem{ref18} M. Merleau-Ponty. Phenomenology of Perception. (Trans.) Commercial Press, 2001.
\bibitem{ref19} L. Chen. Embodied Cognition in Musical Performance: Theory and Evidence. Peking University Press, 2021.
\bibitem{ref20} X. Li. "Machinism: Generative AI Reshaping the Paradigm and Subject Morphology of Music Composition in 2025," China Literature and Art Criticism, no. 3, 2026.
\bibitem{ref21} P. H. Lang. Music in Western Civilization. (Trans.) Guangxi Normal University Press, 2014.
\bibitem{ref22} Z. Han. "A Re-discussion of the Basic Problems of Artificial Intelligence Writing and Music Aesthetics," Art of Music (Journal of the Shanghai Conservatory of Music), no. 1, 2020.
\end{thebibliography}
\end{document}